\documentclass[12pt]{iopart}
\usepackage{epsfig}
\usepackage{graphics}
\usepackage{array}
\usepackage{multirow}
\usepackage[usenames,dvipsnames]{color}
\usepackage{iopams}
\usepackage{mathrsfs}
\usepackage{bbm}
\usepackage{setstack}
\usepackage{makecell}

\newcommand{\beqn}{\begin{eqnarray}}
\newcommand{\eeqn}{\end{eqnarray}}

\begin{document}

\title{Classical Hamiltonian Time Crystals -- General  Theory And Simple Examples}

%\vskip 5.0cm
\author{Jin Dai$^1$,  Antti J. Niemi$^{1,2,3}$, Xubiao Peng$^3$}
%\homepage{http://www.folding-protein.org}
\address{$^1$ Nordita, Stockholm University, Roslagstullsbacken 23, SE-106 91 Stockholm, Sweden}
\address{$^2$ Laboratoire de Mathematiques et Physique Theorique
CNRS UMR 6083, F\'ed\'eration Denis Poisson, Universit\'e de Tours,
Parc de Grandmont, F37200, Tours, France}
\address{$^3$  School of Physics, Beijing Institute of Technology, Haidian District, Beijing 100081, People's Republic of China}
\eads{\mailto{djcn1987@outlook.com}, \mailto{Antti.Niemi@su.se}, \mailto{xubiaopeng@gmail.com}}

\begin{abstract}
%There are {\it No-Go} theorems that exclude timecrystalline solutions of Hamilton's equation, in a phase space that is a symplectic manifold. 
We focus on a Hamiltonian  system with a continuous symmetry, and dynamics that takes place on a presymplectic manifold.  We explain how  the  symmetry
can become  spontaneously broken by a time crystal, that we define as the minimum of the available  mechanical 
 free energy that is simultaneously a time dependent solution of Hamilton's equation. The mathematical description of  such a 
 timecrystalline spontaneous symmetry 
 breaking  builds on concepts of  equivariant Morse theory in the space of Hamiltonian flows.  As an example we analyze a general family of timecrystalline 
Hamiltonians that is designed to model polygonal, piecewise linear closed
strings. The vertices correspond to the locations of pointlike interaction centers; the string is akin a chain of atoms, that are
joined  together by covalent bonds,  modeled by the links of the string.   We argue that the  timecrystalline character of the string can be affected
by its topology.  For this we show that a  knotty string is usually more  timecrystalline than a string with no  self-entanglement. 
We also reveal a relation between phase space topology  and the occurrence of  timecrystalline dynamics. For this we show that in the case of
three point particles, the presence of a time crystal can relate to a Dirac monopole that resides in the phase space. 
Our  results propose that physical examples of Hamiltonian time crystals can be realized in terms of closed, knotted molecular rings.

\end{abstract}

%\pacs{%87.15.Cc 05.45.Yv 36.20.Ey}}

\submitto{\NJP Focus  issue on Time Crystals}

\maketitle

%$\bullet$ Intro
%
%$\bullet$ general theory
%
%$\bullet$ example Lie-Poisson
%
%$\bullet$rotation-from-vibration
%
%$\bullet$ Dirac monopole

%\vskip 1.0cm
\section{Introduction}

Time crystals were originally introduced by Wilczek and Shapere  \cite{Wilczek-2012,Shapere-2012,Wilczek-2013,Sacha-2018}.  They proposed that a
time crystal is a minimum energy configuration, that is also time dependent. As a consequence a time crystal breaks  time translation invariance spontaneously, 
in the same manner how an ordinary crystal breaks space translation symmetry. 
Soon afterwards it was 
argued that time crystals can not exist, in the Hamiltonian context \cite{bruno-2013,watabane-2014}. But
recently explicit examples of  Hamiltonian time crystals have been constructed \cite{Dai-2019a,Dai-2019b,Alekseev-2020}.
A general framework has  also been developed \cite{Alekseev-2020}, it  identifies a set of conditions that are 
sufficient for the existence of a Hamiltonian time crystal:
%The existence of a Hamiltonian time crystal relates to the  wider concept of spontaneous symmetry breakdown. 
Time crystalline dynamics can  take place provided 
Hamilton's equation  has symmetries that give rise to conserved Noether charges. A time crystal breaks the symmetry spontaneously, including
time translation symmetry: A time crystal is simultaneously
both a minimum of the energy and a time periodic trajectory,  that is generated by a definite linear combination of the  
conserved charges. For this kind of timecrystalline spontaneous symmetry breaking to occur the phase space needs to have a presymplectic 
structure \cite{Marsden-1999}. The proper mathematical framework  engages  equivariant Morse theory  in the space of closed 
Hamiltonian trajectories \cite{Wasserman-1969,Niemi-1994,Austin-1995,Nicolaescu-2011}.

Here we explain in detail  the origin and character of  timecrystalline Hamiltonian dynamics; the article is largely a survey of our original
work, published in \cite{Dai-2019a,Dai-2019b,Alekseev-2020}.    
We first explain why conserved Noether charges are necessary for a Hamiltonian time crystal to exist. We describe how a time crystal spontaneously breaks the
symmetry group of Noether charges into an abelian subgroup that generates  periodic timecrystalline motion. As an example 
we analyze in detail a general family of Hamiltonians \cite{Dai-2019a} that support  timecrystalline dynamics. 
The Hamiltonians are designed to model the dynamics of piecewise linear,  polygonal strings: For a physical example,  
the timecrystalline Hamiltonian functions we consider
appear often as energy functions in the context of  coarse grained models of molecular chains \cite{Leach-2001}. The  pertinent conserved quantity that gives 
rise to the  timecrystalline dynamics in our Hamiltonian description simply states the geometric actuality, that  the chain forms a closed string.  We then bring up
that a closed ring is an elemental example of a string with a knot;  a simple closed string is known as the  unknot.  This motivates us to consider 
more complex entangled structures, and we proceed to show that when  the knottiness of a closed string increases  its timecrystalline qualities are usually
enhanced. As an example we analyze in detail a closed polygonal string that forms a trefoil knot.  

We then inquire about the microscopic origin of timecrystalline dynamics, at the level of
the phase space topology. Our starting point is the  well known result in geometric mechanics that when a deformable body contains at least three 
independently movable components,  its vibrational and rotational motions are no longer 
separable \cite{Guichardet-1984,Shapere-1989a,Shapere-1989b,Littlejohn-1997,Marsden-1997}.  Even with 
no net angular momentum, small  local vibrations can cause a global rotation of the entire body.  
We argue that this relation between vibrations
and rotations can provide an explanation of effective timecrystalline dynamics, in the case of a molecular ring. Thus our results  
propose that ring molecules, and in particular those that support a  knot,  are good candidates for 
actual physical examples of Hamiltonian time crystals.

%Finally, we search for a cause of timecrystalline dynamics, from the topological structure of the presymplectic manifold. We argue that, in 
%the actual case of physical molecular chains, the models that we have considered describe effective, collective large time scale dynamics
%of atomic level motions. 
%properties of very 
%examples,    
%Knotted molecules provide  unique opportunities to construct entirely 
%new materials with exceptional strength and elasticity. They have 
%many remarkable  physical and chemical properties, from selective ion binding 
%and strong catalytic activity to generators of molecular machines \cite{Dietrich-1989,Lukin-2005,Sauvage-2008,Erbas-2015,Marcos-2016,Horner-2016,Fielden-2017,Segawa-2019}. 
%Indeed, it  has already been reported that in certain circumstances a knotted molecule appears to undergo autonomous swirling motion, resembling  a 
%molecular motor  \cite{Segawa-2019,Gruziel-2018}. 
%

\section{Theory of Hamiltonian time crystals}

%In this Section we explain how to construct Hamiltonian time crystals in general, following \cite{ }: 
Initially it was thought that  there can not be any energy conserving, Hamiltonian time crystals \cite{bruno-2013,watabane-2014}. 
This is  the conclusion that one arrives at, when one looks at the  textbook Hamilton's equation
\begin{equation}
\eqalign{ \frac{dq^a}{dt} = \{ q^a , H \} = \frac{\partial H}{\partial p^a} \cr
 \frac{dp^a}{dt} = \{ p^a , H \} =  - \frac{\partial H}{\partial q^a}} 
\label{pqH}
\end{equation}
Suppose that ($q^a,p^a$) are (possibly local) coordinates on a phase space that is  a compact closed manifold. Then a minimum of the Hamiltonian energy
function $H(q,p)$ is also its  critical point, so that at the energy minimum the right hand sides of (\ref{pqH}) vanish. As a consequence the left hand sides must 
also vanish, and we immediately conclude that a minimum of $H$ can only be a  time independent solution of Hamilton's equation. In particular, we conclude that
there are no
Hamiltonian time crystals. 

However, there is a way to go around this argument, and we now explain how it goes.  We start with Hamilton's equation that is defined on a $2N$ 
dimensional  phase space which is a symplectic manifold $\mathcal M$; for a background on geometric mechanics see for example  \cite{Marsden-1999}.  
On a symplectic manifold there is always a closed and non-degenerate two-form  $\Omega$,
\begin{equation}
\eqalign{ &  \Omega = \Omega_{ab} d\phi^a \wedge d\phi^b \cr
 d & \Omega = 0
 } 
\label{Omega}
\end{equation}
The $\phi^a$ ($a=1,...,2N$) are  generic local coordinates on the manifold $\mathcal M$. Hamilton's equation is
\begin{equation}
\Omega_{ab} \frac{ d\phi^b}{dt} = \partial_a H
\label{phiH0}
\end{equation}
where the Hamiltonian $H(\phi)$ is a smooth real valued function that is supported by $\mathcal M$.  A solution of Hamilton's equation (\ref{phiH0})  describes 
a trajectory $\phi^a(t)$ on the manifold $\mathcal M$.  The trajectories are non-intersecting,  they are uniquely defined by the
initial values $\phi^a(0)$.
The  inverse  of the matrix $\Omega_{ab}$ determines the Poisson brackets on $\mathcal M$,
\begin{equation}
\{ \phi^a , \phi^b \} = \Omega^{ab}(\phi)
\label{PBphi}
\end{equation}
and we can use the Poisson brackets to  write  Hamilton's equation (\ref{phiH0}) as follows,
\begin{equation}
\frac{ d\phi^a}{dt} \ = \ \{ \phi^a , H \} \ = \ \Omega^{ab} \partial_b H
\label{phiH}
\end{equation}
A  time crystal would be a solution of Hamilton's equation (\ref{phiH})  that has  both a non-trivial $t$-dependence,  and is also a minimum
of the Hamiltonian energy function $H(\phi)$. We now show that despite the {\it No-Go} argument of (\ref{pqH}),  such Hamiltonian time crystals 
do exist provided a set of conditions is satisfied.

For clarity we shall only search for genuine time crystals, those that are periodic functions of time
$\phi^a(t+T) = \phi^a(t)$ for some finite non-vanishing $T$.  Thus a time crystal would spontaneously break 
continuous time translation symmetry into a discrete group of time translations. But we note that depending on $H$ there might also be timecrystalline solutions
that are non-periodic in $T$. They also break time translation symmetry, their properties can be analyzed similarly.

Darboux theorem states that on a symplectic manifold we can always find  a  local coordinate transformation  that sends  
the $\phi^a$ to the canonical momenta and coordinates  ($q^a,p^a$) with their standard 
canonical Poisson brackets. In such Darboux coordinates Hamilton's equation acquires the textbook form (\ref{pqH}) and
without any  additional input, the simple {\it No-Go} argument that is based on (\ref{pqH}) is valid and excludes
timecrystalline solutions of (\ref{phiH0}). 
Thus, to construct timecrystalline Hamiltonian dynamics,  we need to proceed beyond plain symplectic geometry. Such a more 
general framework is {\it presymplectic} geometry.    A presymplectic manifold is 
simply a manifold with a closed two-form. Presymplectic structure  can be encountered, even in a symplectic context,  
for example when  Hamilton's equation is subject to contraints, or when it supports 
continuous symmetries.  On a presymplectic manifold the {\it No-Go} theorems \cite{bruno-2013,watabane-2014} no longer need to be applicable.
Time crystals can exist, and we now explain how this can take place, in the case of Hamilton's equation with continuous symmetries \cite{Alekseev-2020}.

% Let $H(\phi)$ be our Hamiltonian function with a continuous symmetry. 
We start with a Hamiltonian function $H(\phi)$ that describes
dynamics on a $2N$ dimensional 
{\it symplectic} manifold $\mathcal M$; our presymplectic structure emerges  in the context of standard symplectic geometry. 
Thus we start with a non-singular symplectic two-form with components $\Omega_{ab}$. Its inverse matrix $\Omega^{ab}$
determines the Poisson bracket  (\ref{PBphi}) that gives rise to Hamilton's equation (\ref{phiH}). 

We now assume that,  in addition, the Hamiltonian has a continuous symmetry. According to 
Noether's theorem a continuous symmetry  gives rise  to a conservation law.   We denote the 
ensuing conserved charges $G_i(\phi)$ with $i=1,...,n \leq N$.
The Poisson brackets of the conserved charges with the Hamiltonian vanish,
 \begin{equation}
\frac{d G_i}{dt}  \  = \ \{ H, G_i\} = \Omega^{ab} \partial_a H \partial_b G_i =    0
 \label{HG-bra}
 \end{equation}
The  Poisson brackets of the $G_i$ closes, and coincides with the Lie algebra of the symmetry group,
\begin{equation}
\{ G_i, G_j\} = {f_{ij}}^k G_k
\label{LieG}
\end{equation}
We assume that there is no spontaneous symmetry breaking, in the usual fashion. 
Instead we proceed to describe how the symmetry becomes spontaneously broken, by a time crystal.

We introduce the numerical values  of the conserved charges 
\begin{equation}
G_i \left( \phi(0)\right) = g_i
\label{G-val}
\end{equation}
They are specified  by the initial conditions 
$\phi^a(0)$ of Hamilton's equation.  The preimages of $g_i$   foliate the  symplectic manifold $\mathcal M$ into leaves  that are specified 
by the conditions
\begin{equation}
\mathcal G_i^g (\phi) \ = \ G_i(\phi) - g_i = 0
\label{Gg}
\end{equation}
Each regular value of $g_i$  defines a submanifold $\mathcal M_g$ of  the symplectic manifold $\mathcal M$. 
Notably the Poisson brackets of the  (\ref{Gg})  do not close, but instead we obtain
\begin{equation}
\{ \mathcal G^g_i , \mathcal G^g_j \} = {f_{ij}}^k \, \mathcal G^g_k + {f_{ij}}^k g_k
\label{G+g}
\end{equation}
where the right hand side defines a $n \times n$ matrix 
\begin{equation}
\gamma_{ij} (g) = {f_{ij}}^k g_k
\label{gamma}
\end{equation}
In general this matrix is singular. We assume that its  image has a dimension $s\leq n$;  its
kernel then has a dimension  $(n-s)$.  In general these dimensions depend on the numerical
values $\{g_i\}$. 

For given,  fixed values $g_i$ we restrict the non-degenerate symplectic two-form $\Omega$ of $\mathcal M$  to the corresponding  
submanifold $\mathcal M_g$ and we denote the restriction  
by
\begin{equation}
\Omega_{|_{\mathcal M_g}} \ \equiv \  \omega^g = \omega^g(\phi)_{ab} d\phi^a \wedge d\phi^b
\label{ome}
\end{equation}
The two-form $\omega^g$  is closed, but in general the ensuing matrix
$ \omega^g_{ab}$ is degenerate.  Its kernel has dimension ($n-s$) {\it i.e.} the kernel dimension is equal to that of  the matrix (\ref{gamma}). Thus,
whenever $n-s \not= 0$ the submanifold $\mathcal M_g$ that we equip  with the closed two-form (\ref{ome}), 
is  not a symplectic manifold  but a presymplectic manifold \cite{Marsden-1999}.

We shall assume that  the  physical system of interest has the property,  that all those values $\{g_i\}$ that describe the actual physical scenario 
always have $n-s \not= 0$. The two-form $\omega^g$ then determines Hamiltonian dynamics that takes place on a presymplectic submanifold $\mathcal M_g$ of the initial
symplectic manifold $\mathcal M$.  
Since the {\em No-Go} theorem \cite{bruno-2013,watabane-2014} assumes that the Hamiltonian dynamics takes place  
on a symplectic manifold,  it no longer applies to dynamics that takes place on $\mathcal M_g$ 
and we can start searching for a timecrystalline solution of Hamilton's equation. 

We first need to locate the minimum value of the Hamiltonian $H(\phi)$ on the submanifolds $\mathcal M_g$.  
For this we use the method of Lagrange multipliers: 
We introduce $n$ Lagrange multipliers $\lambda^i$ and we extend the Hamiltonian $H(\phi)$ as follows,
\begin{equation}
H \to H_\lambda = H + \lambda^i (G_i - g_i)
\label{crit} 
\end{equation}
The  Lagrange multiplier theorem  \cite{Marsden-1999} states that on a  submanifold $\mathcal M_g$ the minimum  value $\phi^a_{cr}$ 
of the Hamiltonian $H(\phi)$ coincides  with  a critical point ($\phi^a_{cr}, \lambda_{cr}^i$) of the extended Hamiltonian $H_\lambda(\phi)$. 
Thus we can locate  the minimum value of $H(\phi)$ on $\mathcal M_g$  by solving the equations
\begin{equation}
\left\{ 
\eqalign{ \frac{\partial H}{\partial \phi^a}_{| \phi_{cr} } &  =   - \lambda_{cr}^i \frac{ \partial G_i }{\partial \phi^a}_{| \phi_{cr} } \cr
G_i(\phi_{cr})   & =  g_i
 } 
\right.
\label{Glambda}
\end{equation} 
Accordingly, our search of a time crystal proceeds as follows:

\vskip 0.4cm

\noindent
$\bullet ~$  We first use the equations (\ref{Glambda}) and locate the minimum $\phi^a_{cr}$ of  $H(\phi)$
on $\mathcal M_g$, for all those values $\{ g_i \}$ of the conserved charges  that 
correspond to  the physical scenario that we consider.

\vskip 0.2cm
\noindent
$\bullet ~$  We  then proceed and solve from (\ref{Glambda})  the corresponding values  
$\lambda^i_{cr}$  in terms of $\phi^a_{cr}$. 

\vskip 0.2cm

\noindent 
$\bullet ~$  Whenever $\lambda^i_{cr}(\phi_{cr})  \not=0$ we have
a time crystal: The  minimum energy solution $\phi^a_{cr}$ serves as an initial value to the time crystalline solution of 
Hamilton's equation (\ref{phiH}). Thus, in the case of a time crystal we can use (\ref{Glambda}) to rewrite Hamilton's equation (\ref{phiH})
as follows,
\begin{equation}
\eqalign{ \frac{d\phi^a}{dt} &  =  \
 %\Omega^{ab}\partial_b H  =  
 - \Omega^{ab} \lambda^i_{cr}  \frac{\partial G_i}{\partial \phi^b}  \not=0 \cr
 \phi^a(0)  &  =   \  \phi^a_{cr}
 } 
\label{creq}
\end{equation}

\vskip 0.4cm
\noindent
The  existence of a time crystal is a manifestation of spontaneous symmetry breaking, but in a time dependent context: 
The equation (\ref{creq}) states that a
time crystal  is simply a time dependent  minimal energy symmetry transformation that is
generated by a subgroup of the full symmetry group. This subgroup  is spanned by the following linear combination 
of the conserved Noether charges,
\begin{equation}
G_\lambda(\phi) \equiv \lambda^i_{cr} G_i(\phi)
\label{abelG}
\end{equation}
Accordingly the  time crystal breaks the full symmetry group  
(\ref{LieG}) of the Hamiltonian into the  abelian U(1) subgroup  (\ref{abelG}).  
Note that both $H(\phi)$ and $G_i(\phi)$ 
are by construction $t$-independent along {\it any} Hamiltonian trajectory.  Thus the  Lagrange multipliers $\lambda^i_{cr}$  
are time independent. They depend only on the initial configuration  $\phi^a_{cr}$, as determined by the equation (\ref{Glambda}).

\vskip 0.2cm
We conclude this Section and mention that our construction of Hamiltonian time crystals can be rigorously formulated and analysed using the methods of equivariant Morse theory \cite{Wasserman-1969,Niemi-1994,Austin-1995,Nicolaescu-2011}
in the space of loops on a presymplectic manifold.

\section{Family Of Timecrystalline  Hamiltonians}

%\subsubsection{Reduced Yang-Mills-Chern-Simons Hamiltonian}

As an example  of timecrystalline Hamiltonian dynamics we consider 
a polygonal string. The string  is made of  linear links that connect pointlike interaction centers, that are
located at its $N+1$ vertices including the end points \cite{Dai-2019a}. 

In a physical application the interaction centers can model  atoms.  The links of the string  are then the covalent bonds.  
However, at this point we do not propose to describe any specific material system: In any physical application to a stringlike molecule,  
the Hamiltonian approach that we develop should be interpreted in terms of an {\it effective} theory description. An effective theory
aims to describe a complex physical system in terms of a reduced set of variables. The reduced variables should provide an adequate description
of the physical phenomena,  at length and time scales that are large 
in comparison to the characteristic fundamental  level (atomic) length and time scales.  In many circumstances, when there is a separation of scales, such
an effective theory description that builds on a reduced set of variables can be treated as a conventional dynamical system in its own right. In many cases  the
dynamics can be governed by an energy conserving effective Hamiltonian description, in a useful approximation. 

The vertices of the string have coordinates  $\mathbf x_i$ ($i=1,...,N+1$) and the links are the vectors
\begin{equation} 
\mathbf n_i \ = \ %\frac{
 \mathbf x_{i+1} - \mathbf x_i %}{ | \mathbf r_{i+1} - \mathbf r_i |}  
 \ \ \ \ \ (i=1,...,N)
\label{t}
\end{equation}
%with $\mathbf x_{N+1} = \mathbf x_1$ since we assume that the string is closed. 
The $N$ vectors $\mathbf n_i$ are our dynamical degrees of freedom, and we impose on them the following Lie-Poisson bracket
\begin{equation}
\{ n^a_i , n^b_j \} = \delta_{ij} \epsilon^{abc} n^c_i
\label{n-bra}
\end{equation}
The variables $\mathbf n_i$ together with 
their Lie-Poisson brackets are designed to generate any kind of local motion of the vertices,  
except for stretching and shrinking of the links.  Indeed, since
\[
\{ n^a_i , \mathbf n_j \cdot \mathbf n_j \} = 0
\]
for all $i,j$ the bracket preserves the length of $\mathbf n_i$ independently of the Hamiltonian function. For convenience  we
set all $| \mathbf r_{i+1} - \mathbf r_i | = 1$  in the following.  

We note that a Lie-Poisson bracket simply describes how a Poisson manifold {\it i.e.}  
a manifold that is equipped with a Poisson bracket, foliates into symplectic leaves.
Here, for each link $i$ the Poisson manifold is $\mathbb R^3$ and the leaves are the two-spheres $\mathbb S^2$ with radii 
$r_i=|\mathbf x_{i+1} - \mathbf x_i|$.
We can always introduce local Darboux variables ($p_i,q_i$) with their standard Poisson brackets,  
simply by defining
\begin{equation}
\mathbf n \ = \ \left(
\eqalign{ n^1 \cr n^2 \cr n^3 }\right) \ = \ 
r  \left(
\eqalign{ \cos\varphi \sin\vartheta \cr \sin \varphi \sin\vartheta  \cr \ \ \  \cos\vartheta }\right) 
\label{nang}
\end{equation}
The Lie-Poisson bracket (\ref{n-bra}) then reduces to 
\[
\{ \cos\vartheta , \varphi \} = -\frac{1}{r}
\]
and thus  ($\cos \vartheta , \varphi$)$\sim$($p,q$) are Darboux coordinates, but instead    
of  ($\vartheta_i,\varphi_i$) in the present case of a piecewise linear string we find it more convenient to proceed
 in terms of the Lie-Poisson variables $\mathbf n_i$ due to their immediate geometric interpretation.

The Lie-Poisson bracket (\ref{n-bra})
gives rise to the following Hamiltonian equation
\begin{equation}
\frac{ \partial \mathbf n_i } { \partial t } \ = \ \{ \mathbf n_i , H(\mathbf n) \} \ = \ - \mathbf n_i \times  \frac{ \partial H}{\partial \mathbf n_i} 
\label{Hameq}
\end{equation}
To introduce the conserved charges (\ref{HG-bra}), (\ref{LieG}) and to specify the details of the Hamiltonian {\it a posteriori}, 
we start with the end-to-end distance
\begin{equation}
\mathbf G \ = \ \sum\limits^N_{i =1} \, \mathbf n_i \ = \ \mathbf x_{N+1}-\mathbf x_1
\label{Gdef}
\end{equation}
The components satisfy 
\begin{equation} 
\{ G^a , G^b \} =  \epsilon^{abc} G^c
\label{GGbra}
\end{equation}
Here we focus solely  on Hamiltonians that preserve the end-to-end distance
\begin{equation}
\{ H(\mathbf n) , \mathbf G \} \ = \ 0
\label{HG2}
\end{equation}
Furthermore, in the following we shall always assume that the string is closed so that 
\[
g_i \sim \mathbf x_{N+1}-\mathbf x_1=0
\]
For  (\ref{Gg}) we then have
\begin{equation}
\mathbf G \ = \ \sum\limits^N_{i =1} \, \mathbf n_i  = 0
\label{G=0}
\end{equation}
and the Poisson bracket 
(\ref{G+g}) of the corresponding $\mathcal G^a_i$  closes  and coincides with (\ref{GGbra}). Since the matrix (\ref{gamma}) now vanishes the pertinent kernel 
of $\gamma_{ij}$ is three dimensional and in particular it does not vanish:  In the case of a closed string,  the phase space is presymplectic and we are interested
in the ensuing Hamiltonian dynamics.

We search for a timecrystalline solution of (\ref{Hameq}), (\ref{G=0})
using the relevant equation (\ref{Glambda}). For this
we introduce a Lagrange multiplier ${\boldsymbol \lambda}$ and look for extrema of 
\begin{equation}
H_{\boldsymbol \lambda} = H (\mathbf n) + {\boldsymbol \lambda} \cdot \mathbf G
\label{Hlambda}
\end{equation}
The time evolution of the time crystal (\ref{creq}) is then simply
\begin{equation}
\frac{ \partial \mathbf n_i }{\partial t } = - {\boldsymbol \lambda}_{cr} \times \mathbf n_i
\label{ntime}
\end{equation}
with the initial condition 
\[
\mathbf n_i (t=0) = \mathbf n_{i, cr}
\]
where $\mathbf n_{i, cr}$ is the critical point of  (\ref{Hlambda}) that corresponds to minimal $H(\mathbf n)$ value, in the case of a closed string, and  
\begin{equation}
{\boldsymbol \lambda}_{cr} = - \frac{ \partial H } { \partial \mathbf n_i }_{| \mathbf n_{cr} }
\label{lcrit}
\end{equation}
Thus, whenever (\ref{lcrit}) is nonvanishing we have a time crystal.

The Lie-Poisson bracket makes the search for a fixed point of (\ref{Hlambda}) 
straightforward:  We simply extend Hamilton's equation (\ref{Hameq}) into
\begin{equation}
\frac{ \partial \mathbf n_i}{\partial t}  
= - \mathbf n_i \times  \frac{\partial H_{\boldsymbol \lambda}} {\partial \mathbf n_i}  +  \mu \, \mathbf n_i \times (\mathbf n_i \times  
\frac{\partial H_{\boldsymbol \lambda} }{\partial \mathbf n_i} )
\label{eom2}
\end{equation}
where $\mu>0$ is a parameter; note that (\ref{eom2}) also preserves the bond lengths $\mathbf n_i \cdot \mathbf n_i$.  From this we derive 
\begin{equation}
\frac{d H_{\boldsymbol \lambda}}{dt} = - \frac{\mu}{1+\mu^2}\sum\limits_{i=1}^N \left |  \frac{ d \mathbf n_i }{d t} \right |^2 \ \leq \ 0
\label{gilbert}
\end{equation}
Thus, whenever $\mu >0$ the time evolution of (\ref{gilbert}) proceeds towards decreasing values of  $ H_{\boldsymbol \lambda}$ and 
the flow (\ref{gilbert}) continues until it meets a critical point value ($\mathbf n_{i,cr} \, , {\boldsymbol \lambda}_{cr}$).  

The present observations  can be readily developed into a numerical algorithm that we can employ to systematically  locate the critical point  
($\mathbf n_{i,cr} \, , {\boldsymbol \lambda}_{cr}$) for which the Hamiltonian function $H(\mathbf n)$ has a minimal value.

\section{Existence And Simple Examples}

We now present simple examples; the examples show by explicit construction, that Hamiltonian time crystals do indeed exist \cite{Dai-2019a}. 

Our  first example has a  Hamiltonian function of the form
\begin{equation}
{H_1}  \  = \ - \sum\limits_{i=1}^{N} a_i \,  \mathbf n_i \cdot \mathbf n_{i+1} 
% \ + \ {\boldsymbol{\lambda}} \cdot\!  \sum\limits_{i=1}^{N} \mathbf t_i 
\label{H1}
\end{equation}
Clearly, its Poisson bracket with  (\ref{Gdef}) vanishes.
 We start with $N=3$, with (\ref{G=0}) we have $\mathbf n_4 \equiv \mathbf n_1$ so that the  string is closed and its vertices
$\mathbf x_1, \, \mathbf x_2$ and $\mathbf x_3$ are 
the corners of an equilateral triangle. The energy function can only have one value but its derivatives are nonvanishing, and  
the Lie-Poisson bracket (\ref{n-bra}) gives Hamilton's equation
\begin{equation}
\cases{  \frac{d \mathbf n_1 }{dt}  =  \mathbf n_1 \times (a_1 \mathbf n_2 + a_3 \mathbf n_3) \\
\frac{d\mathbf n_2   }{dt}  = \mathbf n_2 \times (a_2 \mathbf n_3 +  a_1 \mathbf n_1) \\
\frac{d\mathbf n_3   }{dt} =  \mathbf n_3 \times (a_3 \mathbf n_1 + a_2 \mathbf n_2) \\
}
\label{n=3}
\end{equation}
This can be easily solved:

For $a_1=a_2=a_3$ we have only the time independent solution and no time crystal, the solution is  an equilateral triangle at rest.

For generic values of  $a_i$ elementary linear algebra shows that  the right hand sides of equations (\ref{n=3}) can not all vanish
simultaneously. The solution which is unique up to time translation, is an 
equilateral timecrystalline triangle rotating around an axis  that is on the plane of the triangle, goes through its center, and
points to a direction that is determined by the parameters ($a_1,a_2,a_3$) as shown in Figure  \ref{fig-1} a).
%\cite{wilczek-2012,yao-2017,zhang-2017,choi-2017}.)
%
%
%
%               FIGURE 1
%
%
%
\begin{figure}
	%\begin{subfigure}{0.6\textwidth}
		\includegraphics[width=1.0\textwidth]{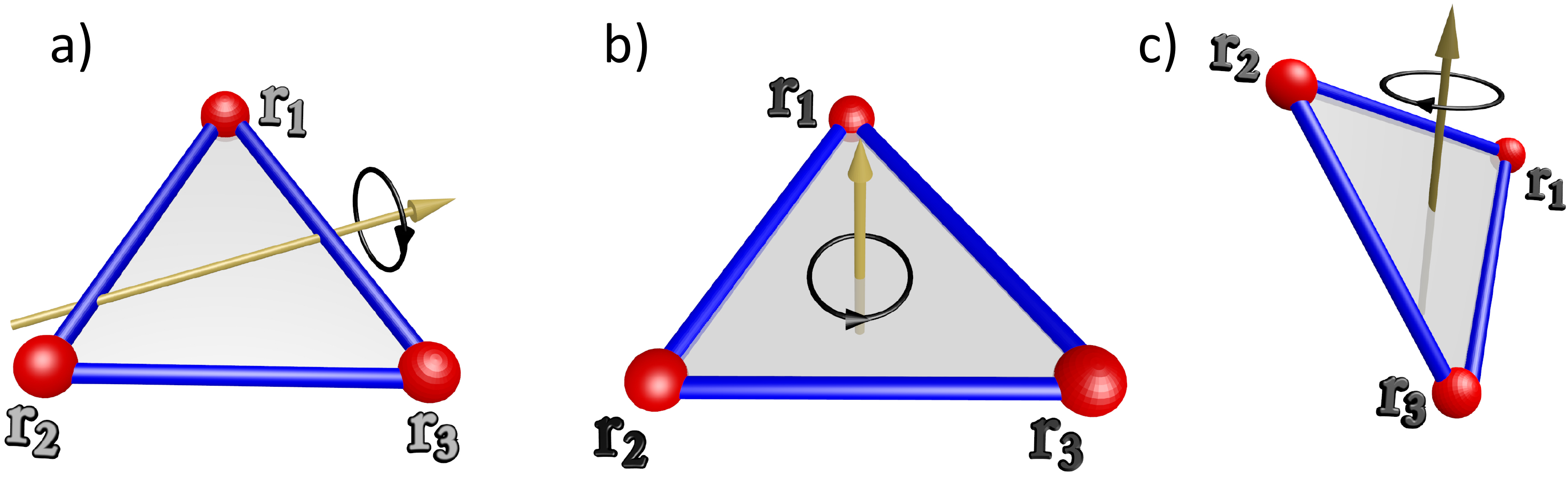}
 		\label{fig-1} 
 		\caption{Figure  a) For generic parameter values ($a_1,a_2,a_3$) the timecrystalline solution of equation (\ref{n=3})
		describes an equilateral triangle rotating around an axis on its plane and thru its center, with direction determined by the parameters. 
		Figure b) For $N=3$ the time crystal of Hamiltonian (\ref{H2}) rotates around an axis which is
		normal to its plane, the direction of rotation is determined by the sign(s) of $b_i$. Figure c) A linear combination of the Hamiltonians (\ref{H1}), (\ref{H2}) with $N=3$
		gives a equilateral triangular  time crystal that rotates around a generic axis thru its center.}
	%\end{subfigure}
  \hfill
  \label{fig-1}
\end{figure}

As another example, the following Hamiltonian
\begin{equation}
{H_2} \ = \
- \sum\limits_{i=2}^{N} b_i  \, \mathbf n_i \cdot (\mathbf n_{i-1} \times \mathbf n_{i+1}) 
% \ + \  {\boldsymbol{\lambda}}  \cdot  \!\sum\limits_{i=1}^{N} \mathbf t_i 
\label{H2}
\end{equation}
also obeys  (\ref{HG2}), and for $N=3$ (so that $\mathbf n_4 \equiv \mathbf n_1$) it supports a time crystal as an equilateral triangle.This time crystal
rotates around its symmetric normal axis as shown in Figure \ref{fig-1}  b).   

The  $N=3$ linear superposition  $H_1 + H_2$ supports  a Hamiltonian time crystal that 
rotates around a generic axis which passes through the geometric center of the equilateral triangle;  the direction of 
the rotation axis and  the speed and orientation of the rotation are determined by
the parameters. See Figure \ref{fig-1}   c).

The present  simple examples prove the existence of Hamiltonian time crystals. But we did not need to 
explicitly search for the critical points of the Hamiltonian (\ref{Hlambda});
when $N=3$ the condition  (\ref{G=0}) can only be satisfied with an equilateral triangle. 
For more than $N=3$ vertices we 
first need to locate the minimal energy configuration $\mathbf n_{i,cr}$ and then solve for the Lagrange multiplier ${\boldsymbol \lambda}_{cr}$ in terms of  $\mathbf n_{i,cr}$.
For this we introduce (\ref{eom2}). As an example we consider  the Hamiltonian (\ref{H2}) with $N=4$, with only one  non-vanishing parameter $b_1 = -1$.  
The minimum energy  configuration  maximizes the volume that is subtended by four unit vectors 
such that 
\[
\mathbf G = \mathbf n_1 +  \mathbf n_2 + \mathbf n_3 + \mathbf n_4 = 0
\]
For energy minimum the  vertices are the four corners of a tetragonal disphenoid  \cite{Conway}, its  faces are isosceles triangles with edge lengths in the proportions 
$\sqrt 3 : \sqrt 3 : 2$.  
Tetragonal disphenoid is a remarkable geometric object.  Unlike the regular  tetrahedron it  tesselate spaces, and it can also 
be constructed by simple foldings of A4 standard paper as the side ratios of A4 are $1:\sqrt{2}$ \cite{Conway,Gibb}.  The Figure \ref{fig-2} shows the structure
and depicts its timecrystalline rotation.    
%
%
%
%               FIGURE 2
%
%
%
{\begin{figure}
	%\begin{subfigure}{0.4\textwidth}
		\includegraphics[width=0.6\textwidth]{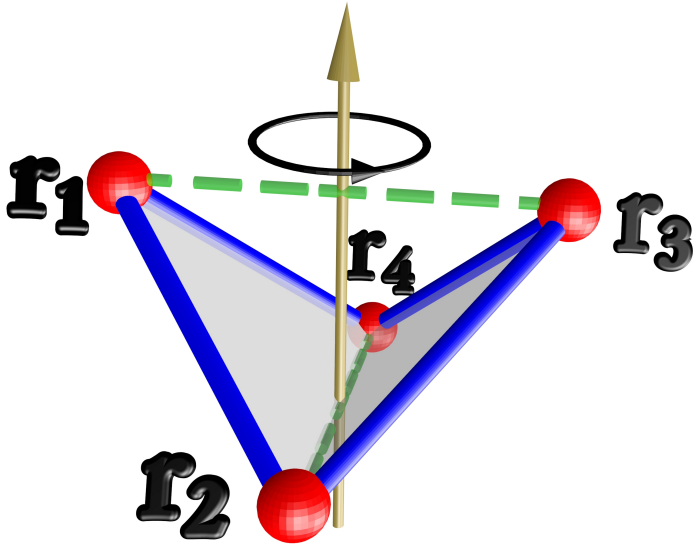}
 		\label{fig-9a}
 		\vskip -0.cm 
 		\caption{ For $N=4$ the time crystal described by the Hamiltonian (\ref{H2})  is a 
		tetragonal disphenoid that rotates around its symmetry axis; the length ratio of the two green segments to the four blue segments is $2 : \sqrt{3}$. The 
		direction of rotation is determined by the sign of $b_i$.}
	%\end{subfigure}
  \hfill
  	\label{fig-2}
\end{figure}

\section{Topology And Time Crystals}

For additional, more elaborate  timecrystalline Hamiltonian functions \cite{Dai-2019b}  we observe that the spatial 
separation $\mathbf x_j - \mathbf x_k$ between any two vertices ($j,k$) along our closed string can always be presented in terms of the
bond vectors $\mathbf n_i$,
\begin{equation}
\mathbf x_k - \mathbf x_j  % \ = \   \mathbf n_j + ... + \mathbf n_{i-1} \ = \  - \mathbf n_{i} - ... - \mathbf n_{j-1} \\
%\la{rn1}
%\end{equation}
%\begin{equation}
=  \frac{1}{2} ( \mathbf n_j + ... + \mathbf n_{k-1})   - \frac{1}{2} (\mathbf n_{k} - ... - \mathbf n_{j-1}) 
\label{rn}
\end{equation}
We have here introduced a symmetrization,  to account for the fact that the vertices $\mathbf x_i$ and  $\mathbf x_k$ are connected  in two different ways 
along the closed string.  Consistent with (\ref{HG2}) we can then add to the Hamiltonian any 
two-body interaction $U(|\mathbf x_k - \mathbf x_j |)$.

In the case of molecular modeling  \cite{Leach-2001}  the vertices of our string describe  atoms or small molecules. They are subject to mutual two-body interactions 
including the electromagnetic Coulomb potential and
the Lennard-Jones potential; the latter is a sum of  the attractive van der Waals interaction and the repulsive Pauli exclusion interaction.
In the case of  charged vertices,  at large distances the van der Waals interaction becomes  small in comparison to the Coulomb interaction, and at
short distances  the Pauli repulsion dominates. Thus, for clarity, here we only consider the Coulomb and the Pauli repulsion interactions. Accordingly we introduce
the following contribution to  our timecrystalline Hamiltonian free energy,
\begin{equation}
U(\mathbf x_1,...,\mathbf x_N)  = \frac{1}{2}  \underset{i\not= j}{\sum_{i,j=1}^{N}}  \frac{ e_i e_j  } { |\mathbf x_{i} - \mathbf x_j | }
+ \frac{1}{2} \underset{i\not= j}{\sum_{i,j=1}^{N}}  \left(\frac{ r_{min} } { |\mathbf x_{i} - \mathbf x_j | }\right)^{\!\! 12}
\label{U}
\end{equation} 
Here $e_i$ is the electric charge at the vertex $\mathbf x_i$ and $r_{min} $ characterizes the extent of the Pauli exclusion; the Pauli exclusion 
prevents string self-crossing, in the case of actual molecules covalent bonds do not cross each other.

We note that a Hamiltonian function such as the linear combination of (\ref{H1}), (\ref{H2}), (\ref{U}) commonly appears  in coarse grained molecular modeling \cite{Leach-2001}: 
The contribution (\ref{H1}) resembles the Kratky-Porod {\it i.e.} worm-like-chain free
energy of local string bending \cite{Kratky-1949}, the contribution (\ref{H2}) includes the effects of string twisting, and (\ref{U}) models interactions that are of long distance
along the string.

As an example we consider a molecular ring with $N=12$ vertices. For the energy function we take 
(\ref{U}), with charged pointlike particles at the vertices.  We inquire how does 
the topology of the  string affect its  timecrystalline character.  For this  we compare two different 
string topologies: We take an unknotted ring, with no entanglement, and we take a ring that is tied into a trefoil knot.  
For numerical simulations we choose $e_i=1$  and $r_{min} = 3/4$ in  (\ref{U}). 

In the case of an unknotted string the flow equation (\ref{eom2}) quickly relaxes into a regular planar 
dodecagon. When we set $\mu = 0$ in the dodecagon, we observe no motion:   A regular planar dodecagon 
with charged point particles at its vertices is not a time crystal.

When we tie the ring into a trefoil the situation becomes different: We first construct a representative initial {\it Ansatz} trefoil for 
the flow equation (\ref{eom2}). We start from the continuum  trefoil 
 \begin{equation}
\cases{  %\vspace{0.1cm}
x_1(s) = L \cdot [\, \cos(s) - A \cos(2s)] \\ %\vspace{0.1cm} 
x_2(s) = L \cdot [\,  \sin(s) + A \sin(2s)] \\
x_3(s) =  \pm \, L \cdot [\,  \sqrt{1+A^2}\sin(3s) ]
}  \ \ \ \ \ \ s \in [0,2\pi)
  \label{3foil}
 \end{equation}
Here $L$ and $A$ are parameters, and the choice of sign in $x_3$ 
determines whether the trefoil is left-handed (+) or right-handed (-).  The initial {\it Ansatz } is highly symmetric, for example
each of the three coordinates have an equal radius of gyration value $R_g=  L \sqrt{1+A^2} $.  
To discretize (\ref{3foil}) for $N=12$, we first divide it 
into three segments that all  have an equal parameter length $\Delta s = 2\pi/3$.  We then divide each of these three
segments  into four subsegments, all with an equal length in space for $N=12$ vertices. 
We set  $A=2$ and when we choose $L=0.340$ each segment has a unit length.  The
three space coordinates ($x_1,x_2,x_3$) have the radius of gyration 
\begin{equation}
R_g^{(i)} \ = \ \sqrt{ \frac{1}{N} \sum_{n=1}^N ( x_i(n)- \bar x_i)^2}
\label{Rg}
\end{equation} 
values ($0.722, 0.722, 0.715$); here $\bar x_i$ is the average of the $x_i(n)$.
This is the  initial  trefoil  {\it Ansatz} that we use in the flow equation (\ref{eom2}). But
we have confirmed that our results are independent of the initial structure we use. 

In our example, of Hamiltonian  (\ref{U}) with parameters ($e_i, r_{min}$) = ($1,3/4$), 
the flow (\ref{eom2}) terminates at a prolate trefoil $\mathbf n_{i,cr}$  with radius of gyration 
values ($0.717, 0.717, 0.889$). When we set $\mu=0$ we find that it is 
a time crystal solution of (\ref{ntime}), (\ref{lcrit}) with angular velocity $\omega \approx 1.571$ in our units. 
%Note that we do not account for energy loss due to electromagnetic radiation effects, in the case of a Coulomb interaction. We assume that radiation effects are
%minor, at a time scale that is pertinent to our analysis.
%
In Figure \ref{fig-3}  we  depict this  time crystalline trefoil, and the way it rotates. 
%
%
%
%
%
%
%
%%%%%%%%%%%%%%%%%%%%%%%%%
%
%
%     Figure-5
%
%
%%%%%%%%%%%%%%%%%%%%%%%%%%
%
%
\begin{figure}[h]
        \centering
                \includegraphics[width=0.95\textwidth]{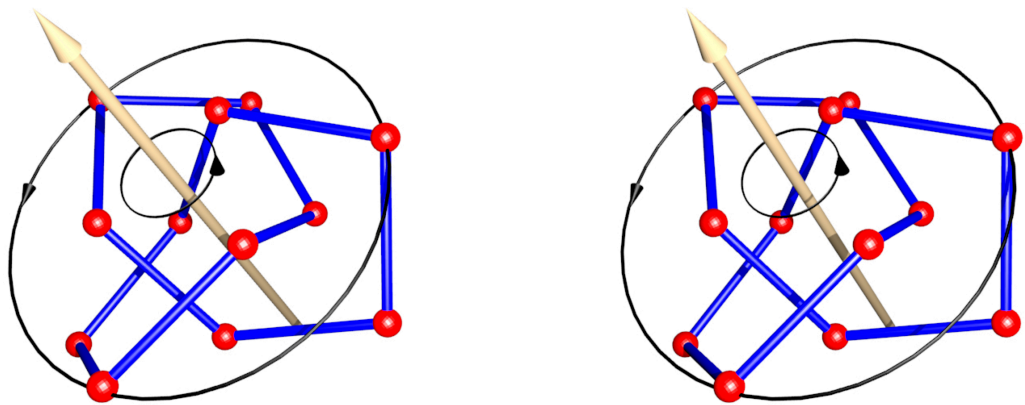}
        \caption{ \small  A 3D cross-eye view of the minimum energy time crystal sol\-u\-tion 
      with Hamiltonian that is a combination of Coulomb attraction and Pauli exclusion.
      The beige arrow is the axis of rotation and the black circles show the direction of rotation. 
      }
       \label{fig-3}
\end{figure}
%
%
%
%
%
%
%
%%%%%%%%%%%%%%%%%%%%%%%%%
%
%
%
%
%
%

The two examples of $N=12$ molecular rings, with the topologies of  an  unknot and a trefoil knot, show a general  relation between
 knottiness  and the  time crystal state that we have observed. The critical point sets $\{ \mathbf x_{i,cr} \}$  of our Hamiltonian functions 
 always pertain to a definite string conformation, with a definite knot
topology.  But  the topology of a generic  knot is in general different from that of the critical point set of the Hamiltonian.   For this reason a knotted molecular
ring often tends to be timecrystalline.

\section{Rotation Without Angular Momentum}

The time crystals that we have constructed are all rotating rigid bodies. Since rotation engages energy and for a time crystal 
that should be minimal, we need to understand the origin of the effective theory
timecrystalline rotational motion:  Why is a rotational motion consistent with minimal mechanical free energy {\it a.k.a.} Hamiltonian.  
For this we first explain how an apparent  rigid rotation can arise in the absence of  any angular momentum, in the case of a 
deformable body.  We then propose that  rotation without angular  momentum can be viewed as the atomic level origin of the time crystals that we have constructed,
in the framework of an effective theory Hamiltonian description.
 
It is well known that when a deformable body contains at least three 
independently movable components,  its vibrational and rotational motions are no longer separable 
\cite{Guichardet-1984,Shapere-1989a,Shapere-1989b,Littlejohn-1997,Marsden-1997}.   Small  local vibrations of a deformable body
can self-organize into a global, uniform  rotation of the entire body.  This is a phenomenon that is  used widely for control purposes.  
For example, the position and altitude of satellites are often controlled by periodic motions of parts of the satellite, such as spinning rotors.

To describe how this kind of atomic level  self-organization  takes place, and how it can lead to an apparent 
 timecrystalline dynamics  at the level of an effective Hamiltonian theory,  
we consider the simplest possible example, that of a 
deformable triangle with three equal unit mass point particles at its
vertices $\mathbf r_i(t)$  ($i=1,2,3$).
We assume that no external forces act on the triangle so that the center of mass  remains stationary,
\[
\mathbf r_1+  \mathbf r_2 + \mathbf r_3 = 0
\] 
at all times.
We also assume that there is no net rotation so  that the total angular momentum $\mathbf L$ vanishes,  
\begin{equation}
\mathbf L \ = \ \mathbf r_1 \wedge \dot {\mathbf r}_1 + \mathbf r_2 
\wedge \dot {\mathbf r}_2 + \mathbf r_3 \wedge \dot {\mathbf r}_3 \ = \ 0
\label{Lz}
\end{equation}
We can orient the triangle to always lay on the $z=0$ plane. 
We then allow the triangle to change its shape in an arbitrary fashion: Two triangles have 
the same shape when they differ from each other only by a rigid rotation, and
we describe shape changes using shape coordinates 
$\mathbf s_i(t)$ with 
\[
\mathbf s_1+ \mathbf s_2 + \mathbf s_3 = 0 
\] 
that we assign to the vertices. These coordinates describe unambiguously 
all possible triangular shapes when we demand that $s_{1x} >0 $, $s_{1y}=0$ and $s_{2y} >0$.

At each time $t$ the shape coordinates  $\mathbf s_i(t)$  relate to 
the space coordinates  $\mathbf r_i(t)$ by a spatial rotation on the $z$-plane,
\begin{equation}
\mathbf r_i(t) \ = \ \mathcal O (t) \mathbf s_i(t) 
\ \ \ \ \ \ {\rm with} \ \ \ \ \ \  \mathcal O (t) \ = \ 
\left( 
\eqalign{
 \cos \theta(t) \  - \sin \theta(t) \\ 
 \sin \theta(t) \ \ \  \ \cos \theta(t) 
 } 
 \right) 
\label{theta}
\end{equation}
We consider a triangle that changes its shape in a periodic, but otherwise arbitrary  fashion; 
the  triangle 
traces a closed loop $\Gamma$ in the space of all possible triangular shapes. We
assume that initially, at time $t=0$, the triangle is {\it e.g.} equilateral and that it
returns back to  its original shape at a later time $t=\mathrm T$. 
During the time period $\mathrm T$ it may have  rotated in space, by an angle  $\theta(\mathrm T)$. 
To evaluate this angle we substitute (\ref{theta}) into (\ref{Lz}). This gives us
\begin{equation}
\theta(\mathrm T)   \ = \ \int\limits_0^{\mathrm T} \! dt \, \, \frac{
\sum\limits_{i=1}^{3}  \left\{ s_{iy} \dot s_{ix}  - s_{ix} \dot s_{iy}\right\}  
}
{ 
\sum\limits_{i=1}^3 \mathbf s^2_i
}
 \ \equiv \ \int_\Gamma d { \mathbf l } \cdot {\mathbf A}  
\label{dotheta}
\end{equation}
We identify here a connection one-form $\mathbf A$, 
it computes the rotation angle $\theta(\mathrm T)$ as a line integral
over the periodic, closed  trajectory $\Gamma$  in the space of all possible triangular shapes 
\cite{Guichardet-1984,Shapere-1989a,Shapere-1989b,Littlejohn-1997,Marsden-1997}.
To  interpret $\mathbf A$ geometrically, we proceed as follows:
We first represent the three coordinates  $\mathbf s_i$  
in terms of  the Jacobi coordinates $\boldsymbol \rho_1, \ 
\boldsymbol \rho_2$  of the classical three-body problem,
 \begin{equation}
\cases{  %\vspace{0.1cm}
\mathbf s_1 = \frac{1}{\sqrt{2}} \boldsymbol \rho_1 - \frac{1}{\sqrt{6}} \boldsymbol \rho_2
\\
%\hspace{-1.2cm} 
\mathbf s_2 = \sqrt{ \frac{2}{3}} \boldsymbol \rho_2 
\\ 
%\hspace{0.3cm}
\mathbf s_3 = - \frac{1}{\sqrt{2}} \boldsymbol \rho_1 - \frac{1}{\sqrt{6}} \boldsymbol \rho_2
}
 \ \ \ \ \ \ s \in [0,2\pi)
\label{jacobi}
 \end{equation}
We denote
\[
\boldsymbol \rho_1     = r \cos\frac{\vartheta}{2}  \left( \eqalign{ \cos\phi _1 \\ \sin\phi_1  }\right)
\ \ \ \ \& \ \ \ \ 
\boldsymbol \rho_2  =  r \sin\frac{\vartheta}{2}  \left( \eqalign{ \cos\phi_2  \\ \sin\phi_2  } \right)
\]
We define $\phi_{\pm} = \phi_1 \pm \phi_2$  and we combine the coordinates into standard spherical coordinates,
\[
\eqalign{
x = r \sin\vartheta\cos \phi_-   
\\
y = r \sin\vartheta \sin \phi_-
\\
z =  r \cos\vartheta
}
\]
The connection  one-form $\mathbf A$ is then 
\begin{equation}
\mathbf A  \ = \  - \frac{1}{2} \cos \vartheta  d\phi_-  -  \frac{1}{2} d\phi_+   \ = \  
\frac{1}{2}  \frac{ \ x d y - y dx } {r (r+z) }  \ - \ \frac{1}{2}  (d\phi_+  + d\phi_-)%  \ \ \ \ (Xubiao ~ says ~ 1/2)
\label{deltatheta}
\end{equation}
where recognize the connection one-form of a single Dirac magnetic monopole in $\mathbb R^3$,  located 
at  the origin $r=0$ and with its  string placed along the negative $z$-axis; see also  \cite{Wilczek-2019}.
Thus the rotation angle $\theta(\mathrm T)$ in (\ref{dotheta}) computes the (magnetic) flux of the Dirac monopole 
through a surface with  boundary $\Gamma$, in the space of all triangular shapes. We note that at 
the location of the monopole all three vertices of
the triangle overlap, and the string corresponds to a shape where two of the vertices overlap. 

We proceed to evaluate the rotation angle  (\ref{dotheta}) in the case of the following (quasi)periodic family of triangles,
\begin{equation}
\mathbf s_1 (t) \ = \  \frac{1}{\sqrt{3}} \left( \eqalign{  \cos ( f[t]) \\  \hspace{0.6cm} 0   }\right) \ \ \ \ \ \ \& \ \ \ \ \ \ \mathbf s_2(t) 
\ = \  \frac{1}{\sqrt{3}} \left(\eqalign{ \cos ( g[t] + \frac{2\pi}{3} ) \\   \hspace{0.6cm} \sin ( %g_2[t]  +
\frac{2\pi}{3} ) } \right)
\label{rot1}
\end{equation}
where we recall that  $\mathbf s_3 (t) = - \mathbf s_1(t) - \mathbf s_2(t)$
and
\[
f(t) = f(t + n\mathrm T_1) \ \ \ \ \  {\rm with } \ \ \ \ \ f(0) = 0
\] 
and 
\[
g(t) = g(t+n\mathrm T_2)\ \ \ \ \ {\rm with} \ \ \ \ \ g(0) = 0
\] 
so that initially at $t=0$ we indeed have an equilateral triangle, with unit length edges. 
We choose
\begin{equation}
f[t] = a  \sin \omega_1 t \ \ \ \  \& \ \ \ \ \ g_1[t] = a \sin  \omega_2 t % \ \ \ \  \& \ \ \ \ \ g_2[t]=0
\label{rot2}
\end{equation}
with amplitude $|a|<1$ and  positive frequencies $\omega_{1,2}$.   

The shape changes (\ref{rot1}), (\ref{rot2}) can be interpreted {\it e.g.} as (small amplitude) oscillations of atoms that are located at the vertices of a triangular molecule. 
We do not specify the mechanism that gives rise to (\ref{rot2}); the origin of these oscillations could be {\it e.g.} quantum mechanical. We are only interested in an effective
theory description,  that becomes valid in the limit of time scales  that are very large in comparison to the periods of the small, rapid vibrational motions.

For generic $\omega_1$ and $\omega_2$  
the integrand of  (\ref{dotheta}) is quasiperiodic, thus by Riemann's lemma we expect that for generic $\omega_{1,2}$  the large 
time limit of $\theta(t)$  vanishes so that there is no net rotational motion.   
However, it turns out that {\it exactly} for $\omega_2 = \pm 2\omega_1 $ the large time limit 
describes a uniformly rotating triangle. To see how this comes about 
we expand the integrand of (\ref{dotheta})  in powers of (small) $a$,  
\[
\frac{ d\theta }{dt} \ = \ - \frac{1}{2} \omega_2 a \cos\omega_2 t  
+ \frac{\sqrt{3}}{12} a^2 \left\{ \omega_2 \sin 2\omega_2 t - \omega_1 \sin 2\omega_1 t \right\} - \frac{1}{8} \omega_2 a^3 \cos 
3\omega_2 t 
\]
\begin{equation}
+ \frac{1}{16} \omega_2 a^3 \left\{ \cos(\omega_2 + 2\omega_1)t + \cos(\omega_2 - 2\omega_1)t \right\} 
+ {\mathcal O}(a^4)
\label{small-a}
\end{equation}
Thus, {\it exactly} when $\omega_2 = \pm 2 \omega_1$ the large time limit of the 
rotation angle $\theta(t)$ increases linearly in time as follows,
\begin{equation}
\theta(t) \ \buildrel {{\rm large}-t} \over {\longrightarrow}  \ \frac{1}{16} \, \omega_2 a^3 t
\label{unirot}
\end{equation}
In Figures \ref{fig-4}  a)-d) we summarize the time evolution of  $\theta(t)$ when we observe its value
stroboscopically, like frames of a movie reel, at regular fixed time steps
$\Delta t(n)  = 10^n$ for $n=0,1,2,4$.
%
%
%
%               FIGURE 4
%
%
%
\begin{figure}
	%\begin{subfigure}{0.6\textwidth}
		\includegraphics[width=0.95\textwidth]{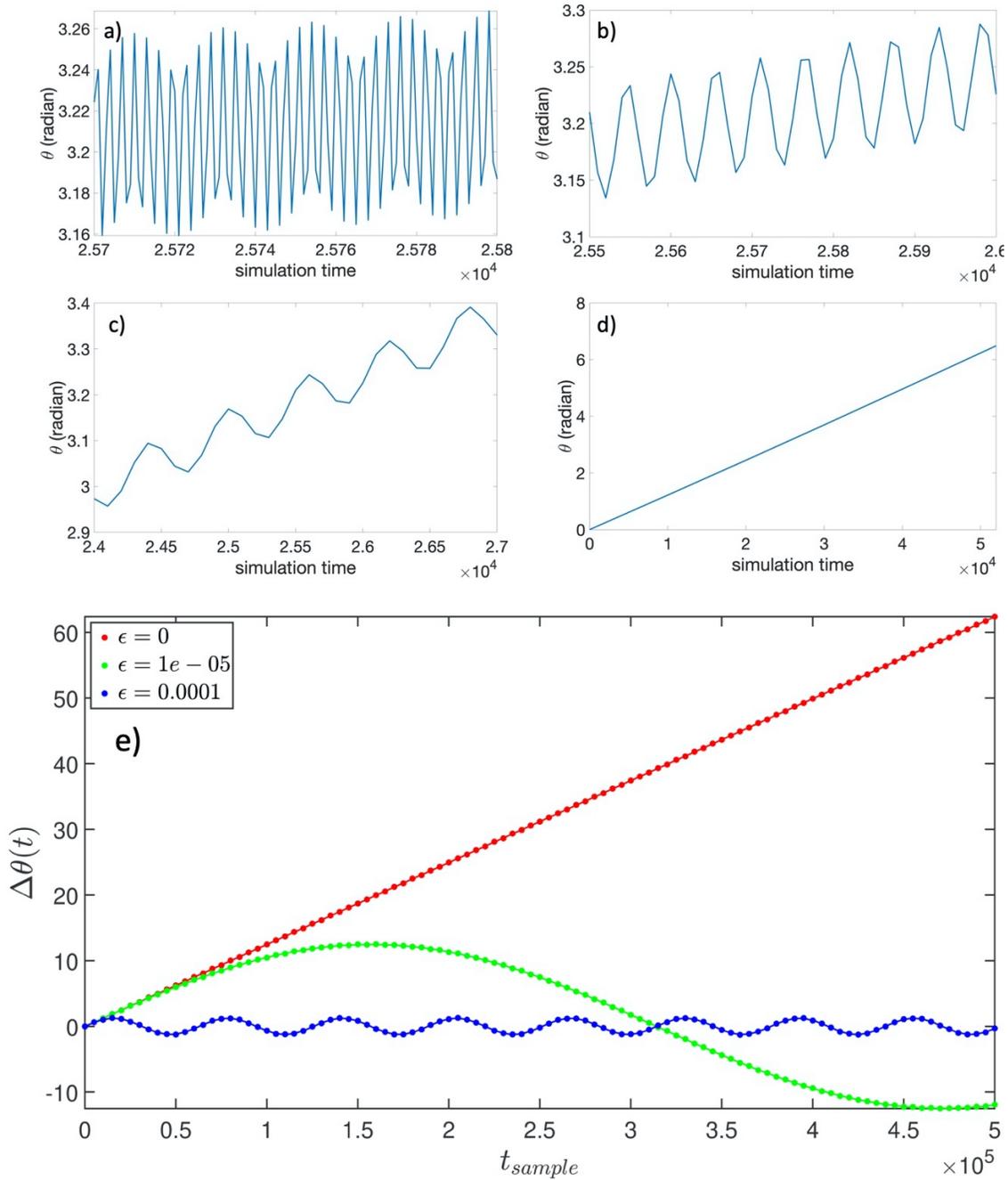}
 		%\label{fig-1}
 		\vskip -0.cm 
 		\caption{Figure a) shows the time evolution of rotation angle $\theta(t)$ in (\ref{dotheta}) when we sample it with 
		stroboscopic time step $\Delta t = 1$.  In Figure b) the time step is increased to $\Delta t = 10$,  
		in Figure c) $\Delta t = 100$ and in Figure d) 
		$\Delta t = 10.000$. In all these Figures  $\omega_2 = 2\omega_1 =2$ and $a=0.1$. 
		Figure e) then shows the transition from uniform rotation at $\omega_2 = 2\omega_1 =2$
		to a sisyphus-like ratcheting motion for $\omega_2=2\omega_1+\epsilon$ with 
		$\epsilon = 10^{-5}$ and $ \epsilon =10^{-4}$.  For each trajectory in Figure e)
		$\omega_1 = 1$, $a=0.2$ and $\Delta t = 10.000$.
		}
	%\end{subfigure}
  \hfill
  \label{fig-4}
\end{figure}

$\bullet~$ The Figure \ref{fig-4} a) shows that when we sample the values of $\theta(t)$ with  
stroboscopic time step $\Delta t = 1$,  the dominant  motion consists of  rapid and slightly irregular   
back and forth oscillations; the irregularities are due to higher order harmonics that are not properly 
caught by the stroboscope.
On top of the rapid oscillations we observe lower frequency undulations, 
five periods are shown in the Figure.  

We also observe a very slow increase in the time averaged 
value of  $\theta(t)$. This indicates the potential presence of a slow clockwise drifting rotation
of the triangle around an axis that is normal to its plane. 

\vskip 0.3cm
$\bullet~$ In Figure \ref{fig-4} b) we increase the stroboscopic  time step to  $\Delta t = 10$. 
We observe slightly irregular back and forth oscillations with an essentially constant amplitude, with a wavelength
that is clearly larger than those  in Figure \ref{fig-4} a). In addition, there  is a much more visible 
increase in the average value 
of $\theta(t)$. It  describes a  
clockwise rotational ratcheting of the  triangle around its  normal axis.

\vskip 0.3cm
$\bullet~$ When the time step increases  to $\Delta t = 100$,  as shown in 
Figure \ref{fig-4} c) the triangle continues to ratchet in the clockwise direction around its  normal axis.
The relative amplitude of the  slightly irregular back and forth oscillations has diminished, while the wavelength has increased.
The same qualitative behaviour persists when we increase $\Delta t = 1000$,  but with increasingly diminished amplitude
and increased wavelength.

\vskip 0.3cm
$\bullet~$ When we increase the stroboscopic time step to the much larger value $\Delta t = 10000$  the motion  
closely resembles that of an equilateral triangle that rotates uniformly around its symmetry axis in clockwise direction, 
with constant angular velocity that can be estimated from  (\ref{unirot}).  See Figure \ref{fig-4} d).

\vskip 0.3cm
$\bullet~$ In Figure \ref{fig-4} e) we show how  the uniform, large stroboscopic time scale  
rotation that we observe for $\omega_2 = 2\omega_1$ and display in  Figure \ref{fig-4} d) 
converts  into back and forth rotations with an amplitude that eventually fades away, 
when $\epsilon = \omega_2 -  2\omega_1$ increases.   

\vskip 0.2cm

From the  Figures we deduce that beyond the back and forth oscillations that dominate
at the very short stroboscopic time scale as shown in Figure \ref{fig-4} a),  there is a transitory  
regime shown in Figures 4 b) and c).   
In this regime the various high frequency
oscillations self-organize into a ratcheting rotation of the triangle. The motion resembles 
the  "Sisyphus dynamics"  described in \cite{Shapere-2012,Shapere-2019}.  Accordingly 
we designate this regime as a pre-timecrystalline sisyphus stage.  We expect that in this 
regime the motion of the triangle 
can be modeled by a version of the ``Sisyphus Lagrangian'' \cite{Shapere-2019}. 

Finally, in the limit of very large stroboscopic time scale shown in Figure \ref{fig-4} d) the
ratcheting motion fades away. In this limit, where we inspect the triangle at time scales that
are much larger than the characteristic timescale of the shape changes {\it e.g.} $T_1 \sim 2\pi/\omega_{1}$,  we can only observe a 
uniform rotation.
In particular, in this  large time scale limit the triangle rotates {\it exactly} in the same manner as the time crystalline triangle 
with Hamiltonian (\ref{H2}) and Poisson bracket (\ref{n-bra}) rotates, as shown in Figure \ref{fig-1} b). 

Thus, we can interpret the Hamiltonian time crystal (\ref{n-bra}), (\ref{H2}) as an effective theory description of the 
deforming $\omega_2 = \pm \omega_1$ triangle, in the large time scale limit.

\section{Summary}

We have identified a  Hamiltonian time crystal  as a time dependent minimum energy symmetry transformation, that  spontaneously 
breaks a continuous symmetry group  into an abelian subgroup. For such a timecrystalline spontaneous 
symmetry breaking to occur, the Hamiltonian dynamics needs to take place on a presymplectic phase space.  

As an example we have analyzed a general family of Hamiltonian  models, designed to describe the dynamics 
of piecewise linear polygonal closed strings. The vertices of the string are pointlike interaction centers, they are 
connected to each other by links that are free to move in any possible way except for stretching, shrinking and chain 
crossing.  

The family of string Hamiltonians that we have analyzed, are commonly encountered in coarse grain descriptions 
of stringlike atoms and small molecules. We have argued that the ensuing timecrystalline Hamiltonian dynamics 
is an effective theory description that becomes valid in a  large  time scale limit, when the very rapid  individual
atomic level  vibrations can be ignored and replaced by much slower collective oscillations. 

Finally, in the case of
a triangular structure, we have found that the effective timecrystalline Hamiltonian dynamics reflects the presence 
of a Dirac monopole in a presymplectic phase space that describes all possible triangular shapes.    

Our results propose that physical realizations of time crystals could be found in terms of  knotted ring molecules.

\section*{Acknowledgements}
JD and AJN thank Frank Wilczek  for numerous discussions. AJN thanks Anton Alekseev for clarifying discussions.
The work by JD and AJN has been supported by the Carl Trygger Foundation, by the 
Swedish Research Council under Contract No. 2018-04411, and by COST Action CA17139. The work by XP is supported 
by Beijing Institute of Technology Research Fund Program for Young Scholars.

%\end{document}

\end{document}